\documentclass[12pt,letterpaper]{article}
\usepackage[utf8]{inputenc}
\usepackage[english]{babel}
\usepackage{amsmath}
\usepackage{amsfonts}
\usepackage{amssymb}
\usepackage{graphicx} 
\usepackage{listings}
\usepackage{color}
\usepackage[left=2cm,right=2cm,top=2cm,bottom=2cm]{geometry}
\usepackage[tagged]{accessibility}
\usepackage[pdflang={en-US}]{hyperref}
\usepackage{xcolor}
\usepackage{float} % Allows putting an [H] in \begin{figure} to specify the exact location of the figure

\allowdisplaybreaks

\newcommand{\authorstyle}[1]{{\large\usefont{OT1}{phv}{b}{n}\color{black}#1}} 

\newcommand{\institution}[1]{{\footnotesize\usefont{OT1}{phv}{m}{sl}\color{black}#1}} 

\usepackage{titling} 

\newcommand{\HorRule}{\color{gray}\rule{\linewidth}{1pt}} 

\pretitle{
	\vspace{-70pt} 
	\HorRule\vspace{10pt} 
	\fontsize{22}{26}\usefont{OT1}{phv}{b}{n}\selectfont 
	\color{black} 
}

\posttitle{\par\vskip 15pt} 

\preauthor{} 

\postauthor{ 
	\vspace{-10pt} 
	\par\HorRule 
	\vspace{20pt} 
}

\title{Using Glowscript to Teach Numerical Modeling in
Undergraduate Biology Education}

\author{
	\authorstyle{Joshua G. Schreibeis, Olivia M. Merideth, and Gavin A. Buxton} 
	\newline\newline 
	\institution{Science Department, Robert Morris University, Moon Township, PA 15108, US.}\\ 
}

\date{}

\begin{document}

\maketitle 

\vspace{-80pt}
Mathematical and numerical modeling is an increasingly important, yet often neglected, topic for biology students. We have found Glowscript to facilitate teaching and introducing computer simulations to students. In particular, the built-in the graphics and graphing capabilities can provide students with immediate feedback. Glowscript is a web-based form of visual Python that runs in a standard browser, and students can easily embed their simulations in external websites. Here we show various examples of how Glowscript is implemented in an undergraduate computational biology course. We hope these examples inspire others to adopt Glowscript in their science classrooms.

\section*{Introduction}

Glowscript (or web VPython) is a simple and versatile programming language for introducing undergraduate biology students to mathematical modeling
and computer simulations. The increasing use of numerical and mathematical modeling in
biological research is not reflected in most undergraduate biology programs
\cite{gross2000education}. Introducing computational biology, and computer programming in
general, to undergraduate biology students can help foster computational thinking, familiarize
students with computer research in their fields of interest, and open up opportunities for
students to use computer models in their future careers \cite{feser2013edge, fawcett2012heavy, buxton2018mathematical}.
However, teaching computer programming to undergraduate biology students must stress the
relevance of computational biology through applications that compliment their other studies and
interests. Here we describe the use of Glowscript for introducing undergraduate biology
students, with no prior programming experience, to computer programming for the first time and to
the use of numerical and mathematical modeling in biological research.\\

Glowscript is a web-based version of visual Python (web VPython) that includes a rich selection of visual objects and a graphing capability that can be used for displaying the results of a
numerical simulation while the simulation is running. Python has been a popular choice for
introducing students to computer programming due to the intuitive syntax \cite{miller2005teaching,
grandell2006complicate, meszarosova2015python, johnson2020analysis}, the ease of defining
variables, and because it is an open and free software \cite{grandell2006complicate,
meszarosova2015python}. That said, Python has been criticized as it overloads operators in a
manner that can be be confusing given dynamic typing, especially when dealing with lists
\cite{johnson2020analysis}, and is not as efficient or optimized as other more syntax-heavy
languages (for example, the use of structures and memory allocation in C can make it much
more efficient). However, Python is a wonderful computer language for introducing students to
computer programming and it forces the programmer to implement a clear indented structure
that facilitates debugging \cite{miller2005teaching, grandell2006complicate}; this is true for both the student and the instructor (who more often than not can be found wandering around the room putting
out fires). This makes the language not only ideal for students to learn, but for teachers to teach.
In addition, the fact that Glowscript offers high-level graphics and graphing capabilities, is
entirely web-based with the students running the code within a standard browser, and that the programs
can be easily exported and embedded in a website, makes Glowscript an invaluable
improvement to the Python language when it comes to both classroom instruction and student adoption outside of class.\\

Here we introduce a couple of different numerical models, implemented using Glowscript, that
can be used in a computational biology classroom; models that capture simple logistic growth,
epidemiology, microtubule dynamic instabilities and drug diffusion from a polymer nanoparticle.
Throughout the implementation of different models in this course, the students are introduced to a basic model, which we will describe in detail here,
and then they are required to creatively adapt or improve the model in some unique way.

\section*{Example: Simple Logistic Growth}

As a simple example, the logistic growth model is a wonderful introduction to using a computer
program to capture a biological process. The logistic growth rate allows for the growth of a
population that is constrained by resources or other limitations of its environment. The model
describes the growth rate of many smaller organisms (such as bacteria, various amoeboid
organisms, and diatoms) reasonably well \cite{vandermeer2010populations}. The population
dynamics are described as
\begin{equation*}
\frac{dN}{dt} = rN\left(\frac{K - N}{K}\right)
\end{equation*}
Where $N$ is the number of organisms, $t$ is time, $r$ is the growth rate and $K$ is the carrying
capacity. The growth of the organism is initially proportional to the number of organisms present,
but will be ultimately limited by the carrying capacity, the number of organisms that the
environment can support. The finite difference approximation of the above equation is given by
\begin{equation*}
N^{t+1} = N^t + \Delta t \, r N^t \left(\frac{K - N^t}{K} \right)
\end{equation*}
Where $\Delta t$ is the time step in the model and the time is indicated through the superscript on $N$.\\

The implementation of this model in Glowscript is shown below.
\begin{lstlisting}
	GlowScript 3.2 VPython
\end{lstlisting}
\vspace{7pt}

The above command loads the glowscript libraries. The constants are created easily in Python,
without the need to first declare the variables or their type.
\begin{lstlisting}
	K = 1000
	r=1
	N0 = 1
	deltat = 0.1
	N = N0
	t = 0.
\end{lstlisting}
\vspace{7pt}

The data can be dynamically plotted, as the simulation is in progress. First, the graph is created
and stored as a variable {\footnotesize\ttfamily plot} in case we want to change any of the properties (which we
currently don’t). The title, axis, and size of the graph are prescribed. Next, the function that we
are plotting is defined as a curve that is red with a width of 2 and labeled ``Population''; even
though there is only going to be one curve, Glowscript will include a key.
\begin{lstlisting}
	plot = graph(title='Logistic Growth.', xtitle='time', ytitle='N', fast=False, width=1000,
																			height=500)
	funct1 = gcurve(color=color.red, width=2, label='Population')
\end{lstlisting}
\vspace{7pt}

The simulation can now be ran, which consists of iterations within a while loop. During each time step the population is updated and plotted as a function of time.
\begin{lstlisting}
	t=0
	while t < 200*deltat:
		rate(500)
		t = t+deltat
		oldN = N
		N = oldN + deltat*r*oldN*(K - oldN)/K
		funct1.plot( t, N )
\end{lstlisting}
\vspace{7pt}

Running the simulation, therefore, dynamically updates the graph of population versus time, and
concludes with the following plot.
\begin{figure}[H] 
\center{\includegraphics[width=\linewidth]{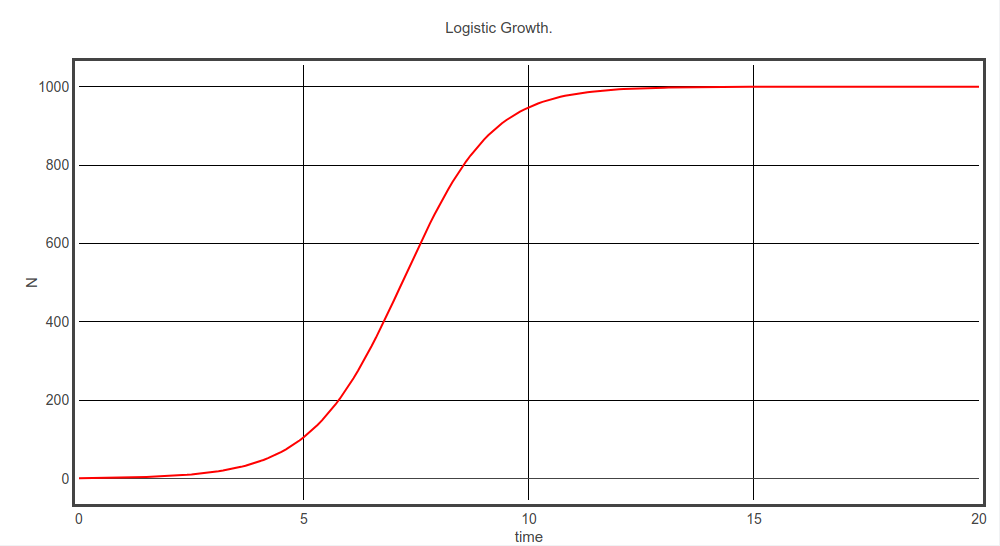}}  
\caption{A screenshot of the Glowscript implementation. The output from the logistic growth model depicts the graphing capabilities of Glowscript. The graph is updated dynamically as the simulation progresses allowing students to visualize the evolution of the system.}\end{figure}
The above logistic growth model, once implemented by the students, can be adapted or
modified in a number of ways. For example, by fitting the above model to experimental data, or
by extending the model to the generalized logistic model, to include more than one competing
population, or to allow for time-varying carrying capacities (commonly used to capture acute
hypoxia in tumors). It is at this point that students are encouraged to be creative, and use the
model to either capture a system of interest or design their own parameter sensitivity analysis.

\section*{Example: Modeling the Zombie Apocalypse}

The course is currently taught during the Fall semester, and in addition to looking at more standard epidemiology models we also look at an epidemiology model of the zombie apocalypse (known as the SIZR model). This model captures the population dynamics of people susceptible to becoming
zombies, the people infected with a zombie virus, the number of zombies and the number of
people removed from the system (dead, but not undead) \cite{munz2009zombies}. The
equation that is used to update the number of susceptible individuals is
\begin{equation*}
\frac{dS}{dt} = \Pi - \beta SZ - \delta S
\end{equation*}
Where $S$ is the number of susceptible people, $t$ is time, $\Pi$ is the birth rate of people, $\beta$ is the rate
at which zombies infect susceptible people, $Z$ is the number of zombies, and $\delta$ is the death
rate of people who are not zombies. The equation for updating the number of infected people is
\begin{equation*}
\frac{dI}{dt} = \beta SZ - \rho I - \delta I
\end{equation*}
Where $\rho$ is the rate at which the infected become zombies (inversely related to the time it
takes to transition). The number of zombies is updated using the following equation
\begin{equation*}
\frac{dZ}{dt} = \rho I - \alpha SZ
\end{equation*}
Where $\alpha$ is the rate at which susceptible individuals are fighting back and destroying the
zombies. Once a zombie is destroyed then it is removed from the system. The number of
removed individuals is of the form
\begin{equation*}
\frac{dR}{dt} = \delta S + \delta I + \alpha SZ
\end{equation*}
Where the susceptible and infected that have died and the zombies killed by susceptible individuals
are included. The above equations can be solved using simple finite difference approximations.\\

The Glowscript code for implementing this model is given below.
\begin{lstlisting}
	GlowScript 3.2 VPython
	Pi = 0.
	beta = 0.1
	delta = 0
	rho = 0.1
	alpha = 0.01
	dt = 0.1
	t = 0.
\end{lstlisting}
\vspace{7pt}

The above code creates and stores variables for some of the constants in the model. The initial
populations are normalized as percentages and the zombies initially comprise just 0.01\% of the
population.
\begin{lstlisting}
	S = 99.99
	I = 0.
	Z = 0.01
	R = 0.0
\end{lstlisting}
\vspace{7pt}

In order to calculate the new values it is necessary to store the old values.
\begin{lstlisting}
	oldS = S
	oldI = I
	oldZ = Z
	oldR = R
\end{lstlisting}
\vspace{7pt}

The syntax for plotting the data starts with declaring a variable that stores the graph, and
includes the title of the graph, the axis and size. Next the different curves that will be plotted are
defined, along with the color of the lines and the labels that will be depicted in the key.
\begin{lstlisting}
	plot = graph(title='Zombie Apocalypse.', xtitle='time (days)', ytitle='percentage', 
											fast=False, width=1000,height=500)
	functS = gcurve(color=color.blue, width=2, label='Susceptible')
	functI = gcurve(color=color.magenta, width=2, label='Infected')
	functZ = gcurve(color=color.red, width=2, label='Zombies')
	functR = gcurve(color=color.black, width=2, label='Removed')
\end{lstlisting}
\vspace{7pt}

We now turn our attention to iteratively solving the simulation. The current implementation consists of 1000 time steps, which are contained in the following while loop.
\begin{lstlisting}
	t=0
	while t < 1000*dt:
		rate(500)
		t = t+dt
		oldS = S
		oldI = I
		oldZ = Z
		oldR = R
\end{lstlisting}
\vspace{7pt}

At each time step, the populations from the previous iteration are stored as the old values and
these old values are used in the finite difference approximation when updating the new values.
\begin{lstlisting}
		S = oldS + dt*(Pi - beta*oldS*oldZ - delta*oldS)
		I = oldI + dt*(beta*oldS*oldZ - rho*oldI - delta*oldI)
		Z = oldZ + dt*(rho*oldI - alpha*oldS*oldZ)
		R = oldR + dt*(delta*oldS + delta*oldI + alpha*oldS*oldZ)
\end{lstlisting}
\vspace{7pt}

The above code is simply the finite difference approximation of the equations above. Once the values have been updated, all that remains is to update the plot. This update occurs
every iteration. This effectively creates an animation of the graph evolving dynamically as the simulation progresses.
\begin{lstlisting}
		functS.plot( t, S)
		functI.plot( t, I)
		functZ.plot( t, Z)
		functR.plot( t, R)
\end{lstlisting}
\vspace{7pt}

The final plot is shown below in Fig. 2. The concentrations for the susceptible, infected, zombie, and
removed populations are depicted as a function of time. Given the values used here, the
zombies very quickly overrun the human population in true apocalyptic fashion.
\begin{figure}[H] 
\center{\includegraphics[width=\linewidth]{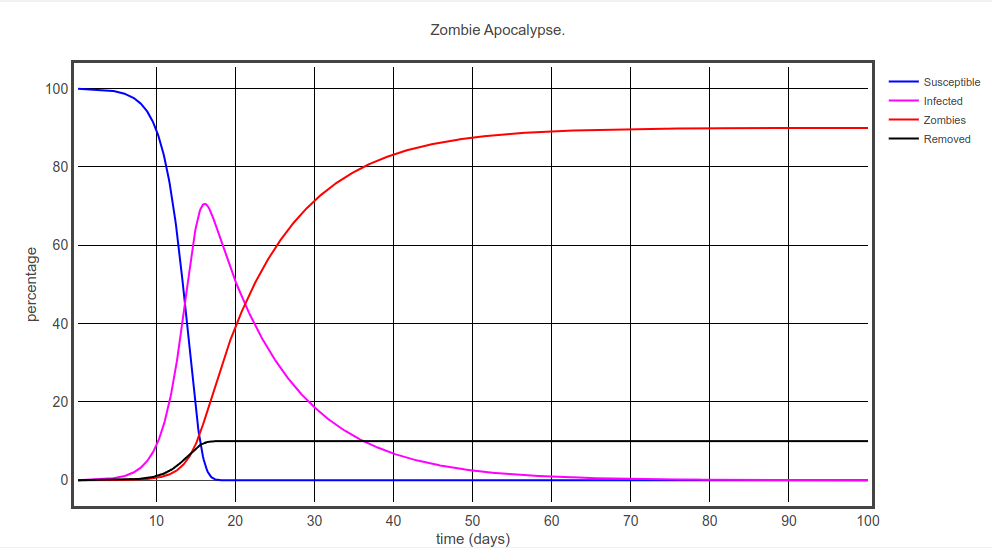}}  
\caption{A screenshot of the Glowscript implementation. The population of susceptible, infected, zombified and removed individuals as a function of time during a zombie apocalypse.}
\end{figure}

Once the basic model is implemented, students can modify the computer program by either
changing the equations or by changing the variables. However, rather than randomly changing
aspects of the numerical model, this system offers a wonderful opportunity to highlight the
connectedness between the mathematical model (and the different terms and parameters in the model) and the system that we are trying to capture. To this end, we
talk about different zombie movies, video games and television shows, and students are asked to write
down physical attributes that they believe describe the zombies in their favorite media. Once we
have these attributes we discuss how to capture these attributes in the model. For example, the
current implementation might not be entirely unreasonable for \emph{Z Nation} or \emph{World War Z}, but
would fail to capture the dynamics of the zombies depicted in the \emph{Night of the Living Dead}, where the humans emerge
victorious at the end. However, the timescale of infection might be too long in this current
implementation for capturing the very fast dynamics of infection in \emph{28 Days Later}. Through the
variation of parameters, or modification of the model, students gain a deeper understanding of
the different terms in the model and how they affect the dynamics of the zombie apocalypse. This is an important illustration for students as to how computer models essentially mimic reality, and how the model can be ``fitted'' to meet our expectations.

\section*{Example: Microtubule Dynamic Instabilities}

In order to highlight some of the graphical capabilities of Glowscript, we consider an example involving microtubule dynamic instabilities \cite{buxton2010mathematical}. Microtubules, as part of the cytoskeleton, provide structure and facilitate cell motility. However, at any given time a subset of microtubules will be growing while others shrink. This allows the microtubules to rapidly explore space, and reorganize if necessary (e.g., during mitosis). To capture microtubule dynamics, various dynamic events in the system have to be accounted for; the nucleation of microtubules, the polymerization of the microtubule (which depends on tubulin and tau concentrations), the depolymerization of microtubules and the hydrolysis of GTP-tubulin in the microtubules to GDP-tubulin (drastically affecting polymerization kinetics). Microtubules are nucleated at a given nucleation site with a rate of $k_{new} = 0.16 \,\text{s}^{-1}$ \cite{piehl2004centrosome}. The GTP-tubulin is converted in the wall of microtubules into GDP-tubulin with a rate of $k_{hydro} = 0.04 \,\text{s}^{-1}$. The polymerization kinetics of a microtubule depends on the local tubulin and tau concentrations and, if the end of the microtubule consists of a GTP-tubulin cap, is taken to be of the form 
\begin{equation*}
k_{poly} = k_{poly,0} \phi_{tub} + k_{poly,\tau} \phi_{tub} \phi_{tau}
\end{equation*}
where $k_{poly,0} = 0.004 \,\text{s}^{-1}\,\mu\text{M}^{-1}$ and $k_{poly, \tau} = 0.01 \,\text{s}^{-1}\,\mu\text{M}^{-2}$ \cite{drechsel1992modulation}. However, if the GTP-tubulin at the end of the microtubule hydrolyzes to GDP-tubulin, then the rate of polymerization is taken to be zero. Similarly, we assume the depolymerization rate, $k_{depoly}$, is 0.7 $\,\text{s}^{-1}$ if the end of the microtubule consists of GDP-tubulin, but is zero if it is a cap of GTP-tubulin \cite{erickson1992microtubule}. Once we have the rates associated with all of the possible events within the system we define an average time over which at least one of these events is likely to occur. In particular, the time step is taken to be the inverse of the sum over all rates.
\begin{equation*}
\Delta t = \frac{1}{\sum_i^N k_i}
\end{equation*}
This time step gives a timescale to the simulation. Assuming an event occurs somewhere in the system, during this time interval, the probability of a particular event occurring is the rate associated with this event relative to the total rate (associated with all events)
\begin{equation*}
P_i = \frac{k_i}{\sum_j^N k_j}
\end{equation*}
where, again, the summation is over all events. An event is chosen to occur from the cumulative distribution function associated with these probabilities and, therefore, the selection of an event takes into consideration the correct probability weightings. The cumulative distribution function is defined as
\begin{equation*}
c_i = \frac{\sum_{j=0}^i k_j}{\sum_{j=0}^N k_j}
\end{equation*}
In order to select an event which will occur during this time step, a random number between 0 and 1 is chosen (RND [0,1]) and the i\textsuperscript{th} event is found for which $c_i < \text{RND [0,1]} < c_{i+1}$. In this manner, the polymerization kinetics of the microtubule dynamics occur stochastically but with the correct probability weightings. \\

Not only do the polymerization rates depend on the local concentrations, but the concentrations are increased or decreased in response to the depolymerization or polymerization of a microtubule. In particular, the concentration of tubulin associated with $0.5 \,\mu\text{m}$ of a microtubule (which we add or subtract in the current model) is $11.6 \,\mu\text{M}$ and the amount of tau is taken to be $0.5 \,\mu\text{M}$. Therefore, the average concentration of tubulin and tau in the cell varies depending on the number and average length of the microtubules. It is worth noting that the local concentrations could also play a role in microtubule dynamics and, for example, a depolymerizing microtubule could leave behind a trail of higher tubulin and tau concentrations which might encourage polymerization along this trail. However, capturing the diffusion of concentrations throughout this three-dimensional system is beyond the constraints of an in-class implementation.\\

One of the wonderful aspects of Glowscript exemplified in this model is the graphics capabilities. In particular, we use curves in Glowscript. This is a function that allows an array of points to characterize a three-dimensional curve through space. Furthermore, each point along the curve can be assigned different thicknesses or colors. For example, here we change the colors of the curves that depict the microtubules in the simulations to show the hydrolysis of GTP-tubulin in the microtubules to GDP-tubulin. Glowscript can include a number of shapes, such as spheres, rectangles and cones (just to name a few). 
\begin{figure}[H] 
\center{\includegraphics[width=\linewidth]{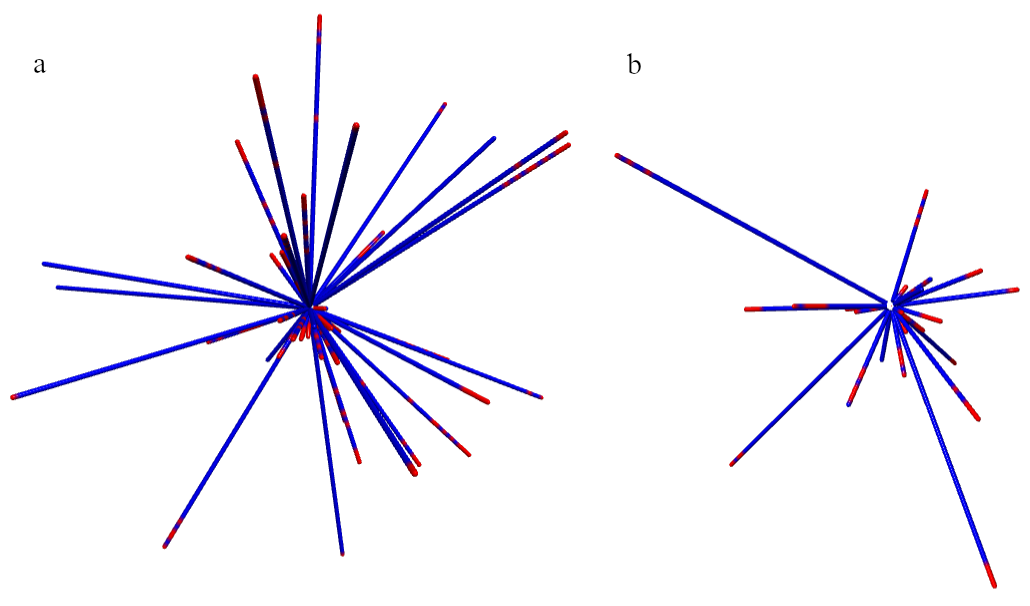}}  
\caption{Screenshot of the Glowscript implementation showing snapshots of the morphology of the microtubules that are displayed as the simulation is running. Systems with a) normal tau concentrations, and b) tau concentrations reduced to 50\%, are contrasted.}
\end{figure}

The output of simulations are depicted in Fig. 3. The microtubules are colored blue if they consist of GDP-tubulin and red if they consist of GTP-tubulin. These simulations took a few minutes to run within a Chrome browser on a typical computer. Fig. 3a shows the morphology of the microtubules for a system with normal levels of tau (around $1 \,\mu\text{M}$).  While Fig. 3b shows the morphology of the microtubules for a system with reduced tau concentrations ($0.5\, \mu\text{M}$). This might mimic a tauopathy where the tau is deposited or aggregated as tau tangles (such as Alzheimer's disease). Clearly a reduction of tau concentrations results in a severe inhibition of microtubule growth. Students can explore this model further by looking at the statistics of the stochastic model. For example, the size distribution of microtubules or the ratio of GDP-tubulin to GTP-tubulin within the microtubules as a function of time. Students can also vary the parameters in the model. For example, as shown here, the initial concentrations of the tau available to the microtubules has a profound effect on the system. therefore, students can investigate the onset of different neurodegenerative diseases through the variations of the parameters in the model \cite{panda2003differential}.

\section*{Example: Drug Encapsulation in Polymer Nanoparticle}

Premedical students, in particular, are often interested in numerical models that capture
therapeutic systems. As an example, we include a model of drug delivery. Polymer
nanoparticles can be used to deliver chemotherapeutic drugs to a tumor, and selectively release
the drug at the tumor site \cite{buxton2012acid, buxton2014simulating}. To introduce the students to these systems we capture the polymer nanoparticle and the diffusion of the encapsulated drug from the
nanoparticle. The polymer can be constructed from acid-labile bonds that disintegrate in
the more acidic environment of the tumor. For computational simplicity (we want this to be solved within a couple of minutes
during class) the system is taken to be spherically symmetric and modeled in one-dimension.\\

In spherical coordinates, the diffusion equation is
\begin{equation*}
\frac{\partial \phi}{\partial t} = \frac{1}{r^2} \frac{\partial}{\partial r} \left( r^2 D \frac{\partial \phi}{\partial r}\right)
\end{equation*}
where $\phi$ is the concentration, $t$ is time, $r$ is the radial distance, and $D$ is the diffusion coefficient. The finite difference approximation is of the form
\begin{equation*}
	\frac{d\phi}{dt} = \begin{cases}
		6\dfrac{D}{(\Delta r)^2}(\phi(1) - \phi(0)) & \text{if } i = 0\vspace{5pt}\\
		0, & \text{if } i = N\vspace{5pt}\\
		\dfrac{D}{(\Delta r)^2i}\left[ (i+1)\phi(i+1) - 2 i \phi(i) + (i-1)\phi(i-1)\right], & \text{otherwise}
	\end{cases}
\end{equation*}
The diffusion of the drug concentration is
\begin{equation*}
	\frac{d\phi_d(i)}{dt} = \begin{cases}
		6\dfrac{D_d}{(\Delta r)^2}(\phi_d(1) - \phi_d(0)) & \text{if } i = 0\vspace{5pt}\\
		0, & \text{if } i = N\vspace{5pt}\\
		\dfrac{D_d}{(\Delta r)^2i}\left[ (i+1)\phi_d(i+1) - 2 i \phi_d(i) + (i-1)\phi_d(i-1)\right], & \text{otherwise}
	\end{cases}
\end{equation*}
where the subscript $d$ indicates quantities associated with the drug. The diffusion coefficient is given by Darcy's law which modifies the diffusion coefficient to account for permeability (here taken to be the space that is not solid polymer).
\begin{equation*}
	D_d(i) = D_{d0} (1 - \phi_s(i))
\end{equation*}
In terms of the nanoparticle we have the solid polymer concentration, $\phi_{ps}$, and the concentration of diffusing polymers that have disintegrated from the nanoparticle, $\phi_p$. The rate at which the solid particle is broken down will depend on the concentration of the blood, and the exposure of the polymer to the more acidic environment near a tumor. The concentration of blood is simply taken to be the volume of the system not occupied by polymer or drug.
\begin{equation*}
	\phi_b(i) = 1 - \phi_d(i) - \phi_p(i) - \phi_{s}(i)
\end{equation*}
The degradation of the solid nanoparticle is then captured using
\begin{equation*}
	\frac{d\phi_s(i)}{dt} = -k \phi_{s}(i)\phi_b(i)
\end{equation*}
where $k$ is a constant, and the degradation depends on the concentration of solid nanoparticle that is being degraded and the concentration of more acidic blood that is responsible for the degradation. As the solid degrades we see an increase in the concentration of the diffusing polymer, whose diffusion can also be captured in spherical coordinates.
\begin{equation*}
	\frac{d\phi_p(i)}{dt} = \begin{cases}
		6\dfrac{D_p}{(\Delta r)^2}(\phi_p(1) - \phi_p(0)) + k \phi_{s}(0)\phi_b(0)& \text{if } i = 0 \vspace{5pt}\\ 
		0, & \text{if } i = N\vspace{5pt}\\
		\dfrac{D_p}{(\Delta r)^2i}\left[ (i+1)\phi_p(i+1) - 2 i \phi_p(i) + (i-1)\phi_p(i-1)\right] + k \phi_{s}(i)\phi_b(i), & \text{otherwise}
	\end{cases}
\end{equation*}
where the diffusion coefficient is again given by Darcy's law
\begin{equation*}
	D_d(i) = D_{d0} (1 - \phi_s(i))
\end{equation*}
Note that the degradation of the polymer nanoparticle could result in a polydispersity in diffusing polymer fragments, and replacing this with a single diffusing polymer concentration is an approximation.\\

The Glowscript implementation (along with all the code discussed here) is provided as
supplementary information. The built-in graphing capabilities allow the concentrations to be
plotted as a function of radial position, and for these plots to animate as the simulation
progresses. Figure 4 depicts an output of a simulation. The solid nanoparticle dissolves and the
polymer fragments are created more at the particle interface and diffuse out. The reduction in
solid nanoparticle concentration increases permeability and the chemotherapeutic drug is able
to diffuse more readily from the nanoparticle.\\

Possible modifications and investigations that students might make to this base model could be
the calculation of drug release rate, the incorporation of different sized polymer fragments (with
different diffusion coefficients), the investigation of the effects of the degradation rate on drug
release rates, the encapsulation of a drug in a hollow nanocapsule, or (in a more advanced
modification) the incorporation of hydrophobic drugs or polymers. Students, therefore, not only implement and use a model of drug delivery from a polymer nanoparticle, but are required to modify and explore this model further. Students, therefore, exhibit creativity and implement scientific methodologies that reflect research practices.
\begin{figure}[H] 
\center{\includegraphics[width=\linewidth]{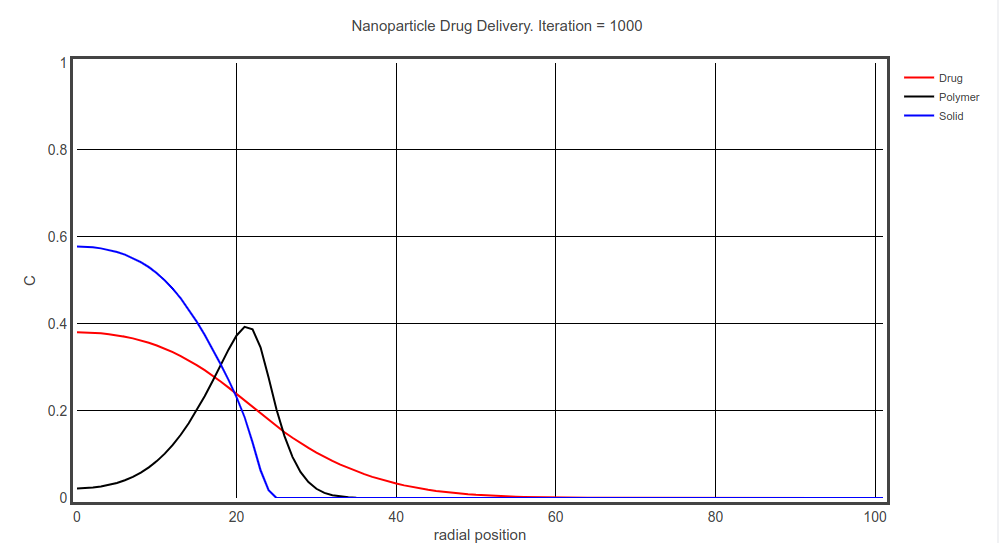}}  
\caption{A screenshot of the Glowscript implementation. The concentrations of solid polymer, diffusing polymer and chemotherapeutic drug from a
nanoparticle.}
\end{figure}

\section*{Conclusions}

Glowscript can be used to introduce computer programming, mathematical modeling and numerical simulations to biology undergraduate students.
Examples of different models that have been implemented in a computational biology course taught at Robert Morris University are highlighted; however, the utility of Glowscript extends beyond the applications considered here.
The ability of Glowscript to display graphical objects, including a dynamic plotting functionality, makes this an ideal programming environment for introducing computer simulations to science students.
One aspect of this course that we have observed over the years is that a number of biology students have been motivated to implement computer models outside of this class, especially in senior research projects. 
We hope that others will read this article and be inspired to implement Glowscript within their classrooms, especially in the biological sciences where students are traditionally less exposed to numerical modeling.

%\section*{References}

\bibliography{paper.bib}
\bibliographystyle{plain}

\end{document}